\documentclass[aps,prd,preprintnumbers,showpacs,twocolumn,groupedaddress,nofootinbib]{revtex4}
\usepackage{graphicx}
\usepackage{natbib}
\usepackage{latexsym}
\def\beq{\begin{equation}}
\def\eel{\end{equation}}
\def\bey{\begin{eqnarray}}
\def\eey{\end{eqnarray}}

\def\sigv{\langle\sigma v\rangle}

\def\lsim{\mathrel{\raise.3ex\hbox{$<$\kern-.75em\lower1ex\hbox{$\sim$}}}}
\def\gsim{\mathrel{\raise.3ex\hbox{$>$\kern-.75em\lower1ex\hbox{$\sim$}}}}

\newcommand{\cm}{{\mathrm{cm}}}
\newcommand{\second}{{\mathrm{s}}}

\begin{document}

\title{Prospects for Detecting Gamma Rays from Annihilating Dark Matter in Dwarf Galaxies in the Era of DES and LSST}

\author{Chen He$^{1}$, Keith Bechtol$^{2}$, Andrew P.~Hearin$^{3}$, Dan Hooper$^{2,3,4}$}
\affiliation{$^1$Department of Physics, University of Chicago, Chicago, IL 60637}
\affiliation{$^2$Kavli Institute for Cosmological Physics, Chicago, IL 60637}
\affiliation{$^3$Center for Particle Astrophysics, Fermi National Accelerator Laboratory, Batavia, IL 60510}
\affiliation{$^4$Department of Astronomy and Astrophysics, University of Chicago, Chicago, IL 60637}

\date{\today}

\begin{abstract}

Among the most stringent constraints on the dark matter annihilation
cross section are those derived from observations of dwarf galaxies by
the Fermi Gamma-Ray Space Telescope. As current (e.g., Dark Energy
Survey, DES) and future (Large Synoptic Survey Telescope, LSST) optical
imaging surveys discover more of the Milky Way's ultra-faint satellite
galaxies, they may increase Fermi's sensitivity to dark matter
annihilations. In this study, we use a semi-analytic model of the
Milky Way's satellite population to predict the characteristics of the
dwarfs likely to be discovered by DES and LSST, and project how these
discoveries will impact Fermi's sensitivity to dark matter. While we
find that modest improvements are likely, the dwarf galaxies
discovered by DES and LSST are unlikely to increase Fermi's
sensitivity by more than a factor of $\sim$2-4. 
However, this outlook may be conservative, given that our model underpredicts the number
of ultra-faint galaxies with large potential annihilation signals actually discovered in the Sloan Digital Sky Survey. 
Our simulation-based
approach focusing on the Milky Way satellite population demographics complements
existing empirically-based estimates.

\end{abstract}

\pacs{95.35.+d, 98.56.Wm; FERMILAB-PUB-13-333-A}
\maketitle

\section{Introduction}

Weakly interacting massive particles (WIMPs) are a leading class of candidates for the dark matter of our universe. In many models, the pair-annihilation of WIMPs can produce potentially observable fluxes of energetic particles, including gamma rays. In recent years, observations of the Milky Way's dwarf spheroidal galaxies by the Fermi Gamma-Ray Space Telescope~\cite{Ackermann:2014sat,aaa+11,GeringerSameth:2011iw} (as well as by ground based gamma-ray telescopes~\cite{Aharonian:2007km,Abramowski:2010aa,Aliu:2012ga,Grube:2012fv,Aleksic:2011jx,Aliu:2008ny}) have yielded constraints on the dark matter annihilation cross section that are among the strongest produced to date, comparable to (and less subject to astrophysical uncertainties than) those derived from observations of the Galactic Center~\cite{Hooper:2012sr} and from searches for dark matter subhalos~\cite{Berlin:2013dva}. For dark matter particles that are lighter than a few tens of GeV and annihilate to quarks, these upper limits are near the canonical cross section predicted for the simplest thermal relics, $\sigv \sim 3\times 10^{-26} \, \cm^3\,\second^{-1}$.

\begin{figure*}[t!]
\includegraphics[width=3.48in]{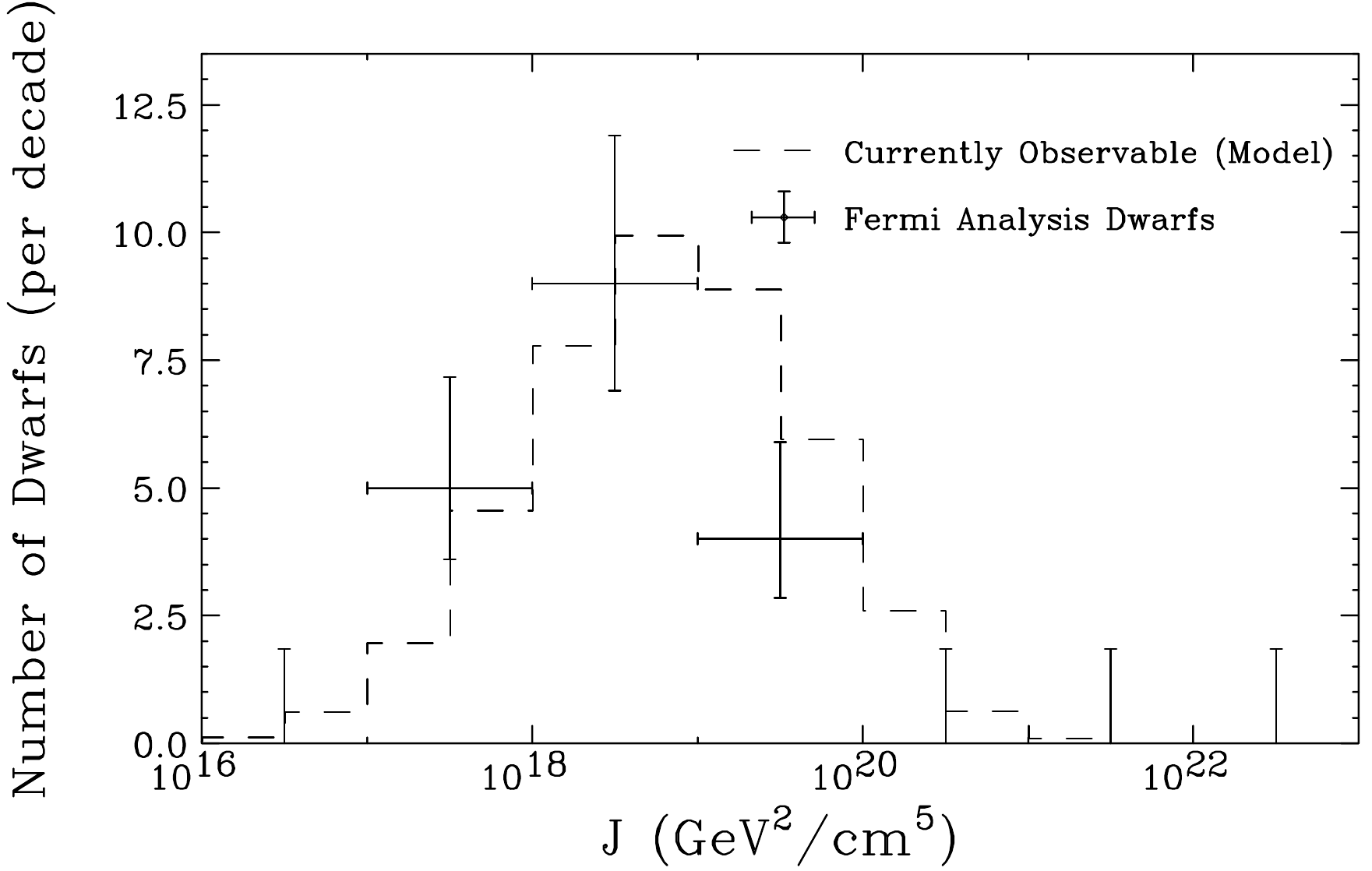}
\hspace{0.02in}
\hspace{-0.3in}
\includegraphics[width=3.48in]{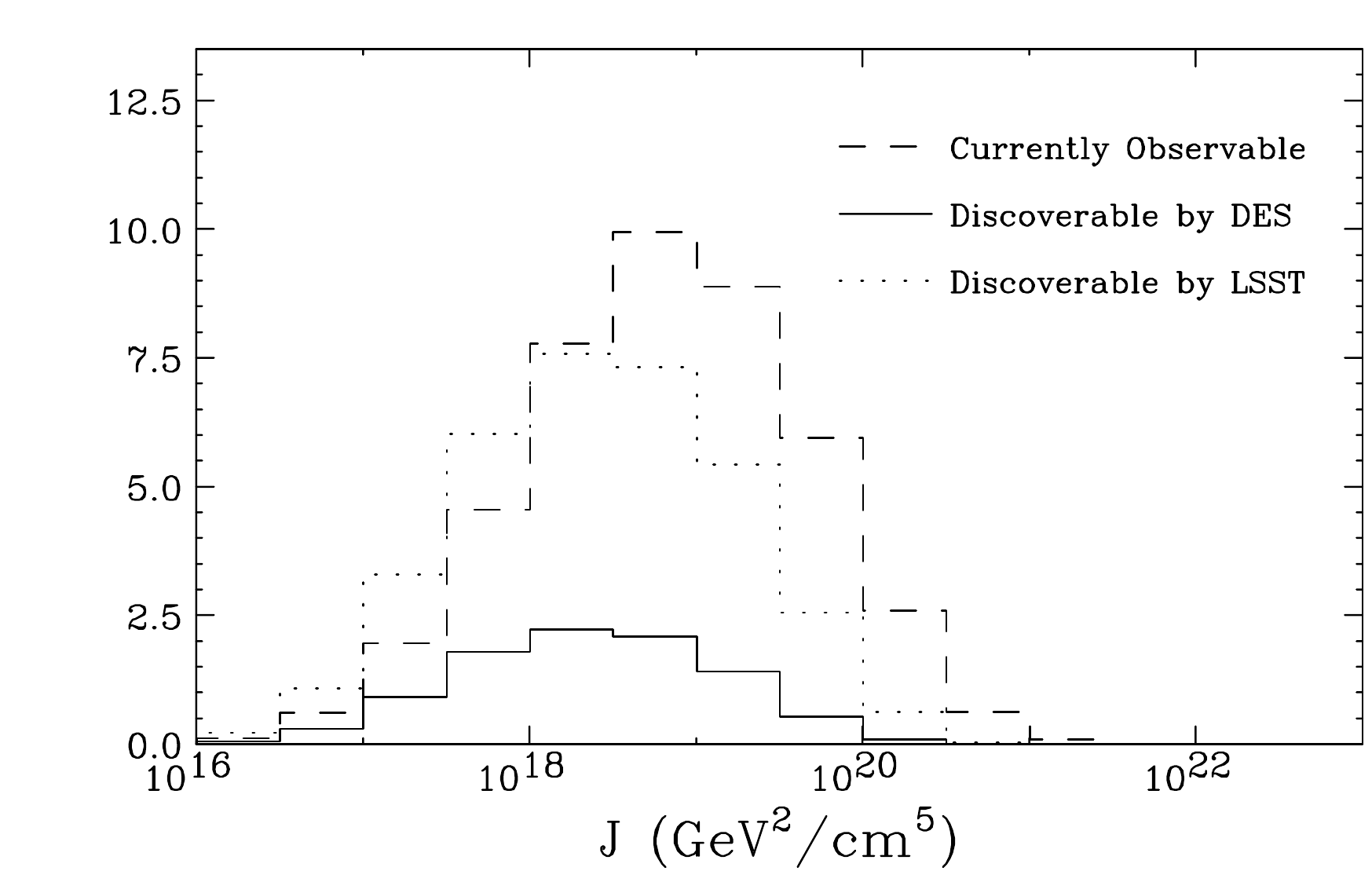}
\caption{Left frame: The distribution of $J$-factors (see Eq.~\ref{Jfactor}) in our model of all currently observable (pre-DES) dwarf galaxies, averaged over 3890 realizations (dashed).  For comparison, we show the distribution of the $J$-factors (with Poisson error bars) of the dwarf galaxies used by the Fermi collaboration in their search for dark matter annihilation products~\cite{Ackermann:2014sat}. Right frame: The distribution of $J$-factors of the dwarf galaxies projected to be discovered by DES (solid) and LSST (dotted). Although the average $J$-factor of DES-discovered dwarfs is expected to be lower than in the currently observable sample (dashed), the tail of this distribution to large $J$-factors is potentially important, and could lead to improvements in Fermi's sensitivity to dark matter annihilations.}
\label{Figure1}
\end{figure*}

Dwarf galaxies are promising targets for indirect dark matter searches
due to their relatively high densities of dark matter and low levels
of astrophysical
backgrounds~\cite{Evans:2003sc,Bergstrom:2005qk,Strigari:2006rd,Strigari:2007at,SanchezConde:2007te,Martinez:2009jh,Kuhlen:2009jv}.
At present, such searches are limited to 16 dwarfs discovered in
the northern hemisphere by the Sloan Digital Sky Survey (SDSS), as
well as 9 previously known classical dwarfs. Future discoveries of
additional dwarf galaxies could improve Fermi's sensitivity to dark
matter, and possibly to a significant degree. In particular, we expect
the currently operating Dark Energy Survey (DES)~\cite{aaa+05} and the
future Large Synoptic Survey Telescope (LSST)~\cite{aaa+09} (scheduled
for 2022), both imaging southern skies, to roughly double the catalog
of known satellite galaxies of the Milky Way~\cite{rsg+11}. In this study, we
forecast the characteristics of the dwarf galaxies within the reach of
DES and LSST, and estimate to what degree Fermi's sensitivity to dark
matter is likely to increase as a result of these forthcoming
discoveries. 

Our focus is on the demographics of the satellite
population, and on the importance of which \textit{particular} new
dwarfs are found, rather than on precisely quantifying the
instrumental sensitivities of Fermi and upcoming optical
surveys. See Ref.~\cite{Drlica-Wagner:2013} for a more empirically-based forecast emphasizing Fermi
analysis methods and assuming that newly discovered satellites
have the same distribution of $J$-factors (defined in Section~\ref{model}) as those discovered by SDSS.

\section{Modeling the Satellite Galaxies of the Milky Way}\label{model}

To model the population of dwarf galaxies within the halo of a Milky Way-like galaxy, we have used the results of Ref.~\cite{fbb+11}, which employed semi-analytic techniques to describe the baryonic physics (including ionization, heating, and cooling of gas, as well as star formation and evolution) relevant to the development of satellite galaxies in the largest subhalos of the Aquarius simulation~\cite{Springel:2008aquarius}. Drawing from this distribution of 505 simulated dwarf galaxy masses and luminosities (kindly provided to us by the authors of Ref.~\cite{fbb+11}), and also adopting a spatial distribution of satellite galaxies derived from Aquarius, we created 3890 realizations of a Milky Way-like system. 

To determine whether a satellite galaxy is detectable in a given
realization, we apply the following criteria. First, we consider any
satellite brighter than $\mathrm{M_v}=-8.9$ (equal to that of the
faintest dwarf discovered prior to SDSS) and well outside of the
Galactic Plane ($|b|>20^{\circ}$) to be a ``classical'' dwarf,
discovered prior to recent surveys.\footnote{This definition
  expressed in terms of optical luminosity, though somewhat arbitrary, cleanly partitions the pre-SDSS and
  SDSS-discovered Milky Way satellites. For comparison, the authors of
  \cite{kleyna:1997} estimated a completeness threshold of $\mathrm{M_v}=-8.8$ to a distance of 180
  kpc based on their systematic search through COSMOS/UKST survey data
  at Galactic latitudes $b<-15^{\circ}$. See Ref. \cite{willman:2010} for a
  review of optical detection limits prior to SDSS.}
For a dwarf to have been detected by SDSS, we require that it resides within the region of the sky covered by the survey, and exceed the magnitude threshold described in Ref.~\cite{wwj09} (following Ref~\cite{wwj09}, we determine by Monte Carlo which near-threshold dwarf galaxies are classified as detectable). Based on the relative thresholds for SDSS~\cite{aaa+08,aaa+07} and DES~\cite{rsg+11,aaa+05}, we adopt a detection criteria for DES which is more sensitive than SDSS by 1.9 in absolute V-band magnitude. Similarly, we adopt a improvement of 5.3 magnitudes in sensitivity for LSST~\cite{aaa+09}. Due to the challenges involved in identifying dwarfs with a large angular extent, we do not consider any dwarfs located within 10 kpc of the Solar System. 
%

To normalize the total number of satellite galaxies in the halo of the Milky Way, we require in each realization that 25 such systems be either discoverable by SDSS or qualify as a classical dwarf (as defined above). On average, we predict that DES and LSST will discover 4.9 and 17.2 previously unknown dwarf galaxies, respectively. At the approximately 1 sigma level, this corresponds to the discovery of 3-8 new dwarfs by DES and 14-21 by LSST.  Note that these ranges reflect only statistical uncertainties, and rely on the validity of our model, as based on the results of Ref.~\cite{fbb+11}.


The flux of gamma rays from dark matter annihilations in a given dwarf galaxy is given by:
\begin{equation}
\Phi \equiv \frac{\sigv \, N_{\gamma}\, J}{8 \pi m^2_{\rm DM}},
\label{flux}
\end{equation}
where $m_{\rm DM}$ is the mass of dark matter particle, $N_{\gamma}$ is the number of gamma rays produced per annihilation (which depends on the mass and annihilation channels of the dark matter particle), and the quantity $J$ encompasses the distribution of dark matter within the dwarf:
\begin{equation}
J \equiv \int_{\Delta \Omega} \int_{l.o.s.} [\rho(l,\psi)]^2 \;dl\; d\Omega,
\label{Jfactor}
\end{equation}
where $\rho$ is the dark matter density and the integral is performed along the line-of-sight. For the solid angle, $\Delta \Omega$, we consider a cone of radius $0.5^{\circ}$, identical to that used in Ref.~\cite{Ackermann:2014sat}. For the dark matter distribution of the subhalos, we assume an NFW profile~\cite{nfw96,Navarro:1996gj} with concentrations given by the mass-concentration relation in Ref.~\cite{mmg+11} using the subhalo mass at the time of accretion.\footnote{Although it has been argued that at least some dwarf spheroidal galaxies may possess dark matter profiles that are shallower than NFW~\cite{Walker:2012td,Walker:2011zu,Walker:2011fs} (see, however, Sec.~4.4 of Ref.~\cite{Strigari:2012gn}), the $J$-factors of such systems are expected to be only modestly impacted by the innermost densities (for example, see discussion in Ref.~\cite{Ackermann:2014sat}).} 
We further take each satellite halo to be tidally stripped beyond a radius determined by the Jacobi limit, as prescribed in Ref.~\cite{BT}, which is applicable when the size of the satellite is much smaller than the distance to the Galactic center.
The effective mass of the Galaxy in this case is the mass interior to the subhalo orbit, assuming an NFW profile for the main halo.
We caution that both tidal heating and tidal stripping may alter subhalo profiles such that the NFW approximation is no longer strictly valid, especially for the innermost subhalos, but a detailed investigation of these effects using numerical simulations is beyond the scope of the current work.


In Fig.~\ref{Figure1}, we plot the distribution of dwarf galaxy
$J$-factors produced by our model, averaged over 3890
realizations. In the left frame, we show the distribution of all
currently detectable dwarfs (classical or detectable by SDSS) and compare this to those of the dwarfs used by the
Fermi collaboration in their search for dark matter annihilation
products~\cite{Ackermann:2014sat}.  From this comparison, we see that our
simulation-based distribution is in good agreement with the
distribution of actual dwarfs studied by Fermi. 


In the right frame of Fig.~\ref{Figure1}, we show the distributions of
dwarfs predicted to be discovered by DES and LSST according to our
model. The populations predicted to be discovered by DES and LSST
exhibit somewhat smaller average $J$-factors in part due to the
ability of these surveys to detect dwarf galaxies at larger distances.


\begin{figure*}[t!]
\includegraphics[width=3.52in]{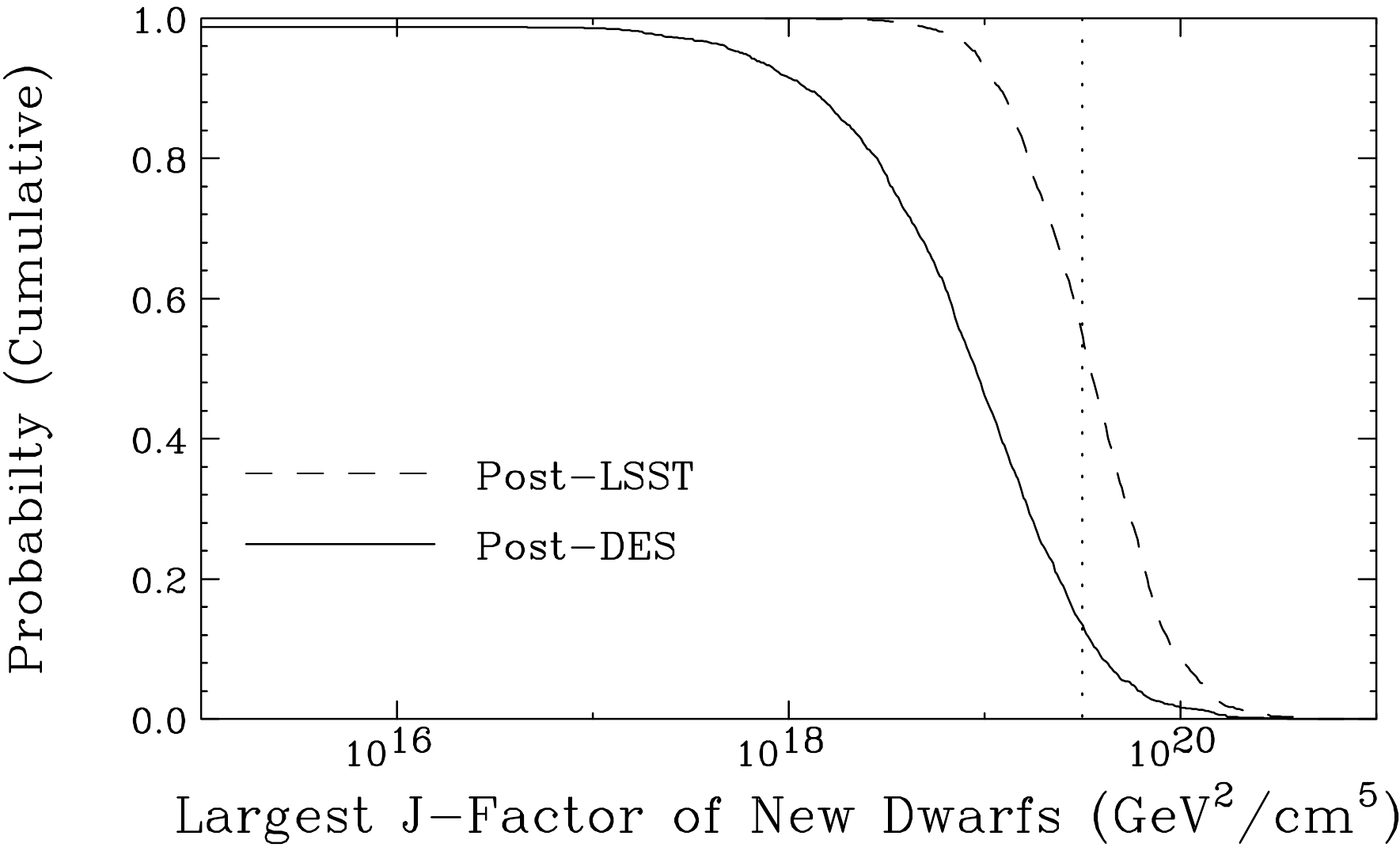}
\hspace{-0.2in}
\includegraphics[width=3.52in]{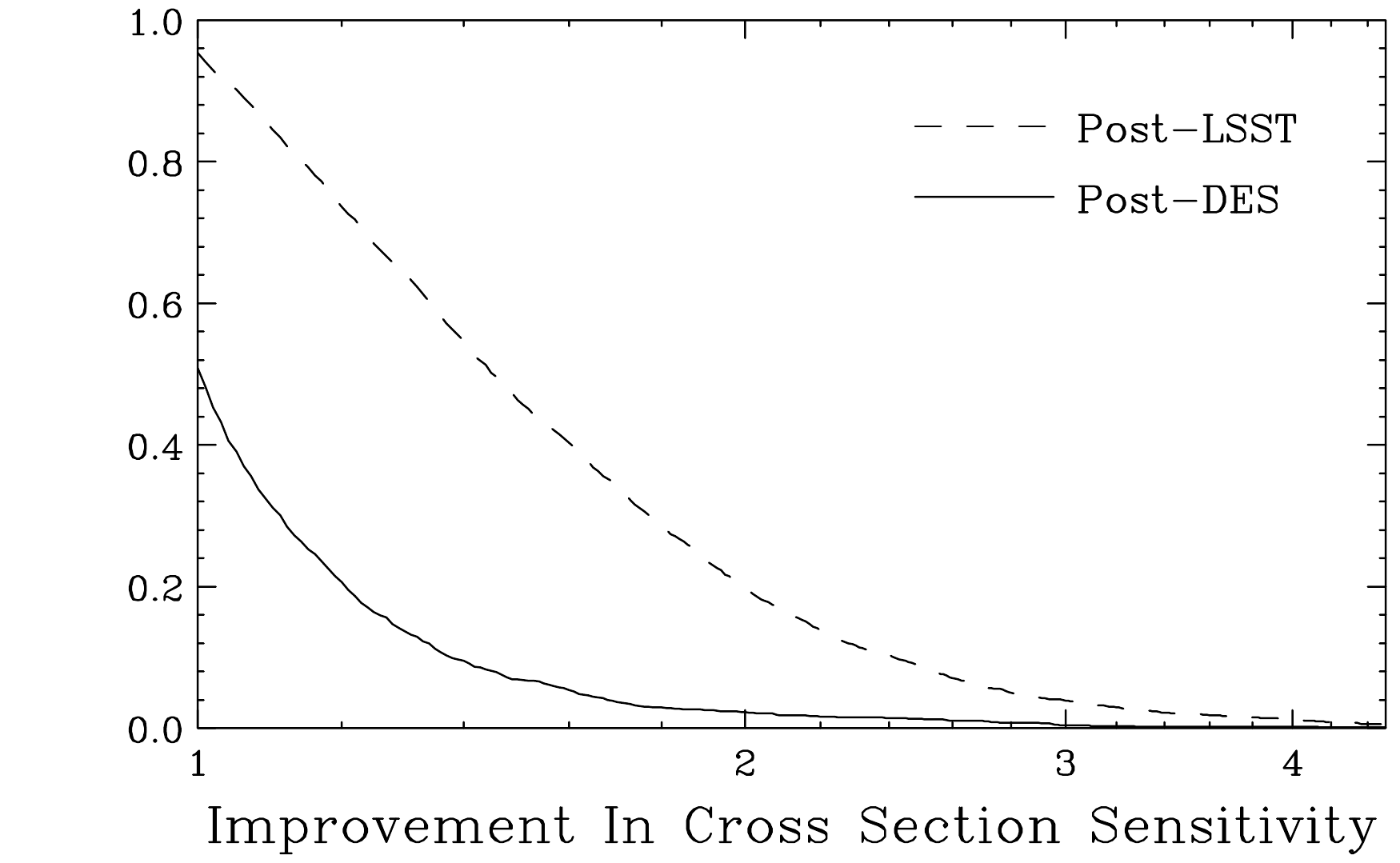}
\caption{Left frame: The estimated probability that DES or LSST will
  discover a dwarf galaxy with a $J$-factor greater than a given
  value. The vertical dotted line denotes the largest $J$-factor of
  the currently known dwarfs ($10^{19.5}$ GeV$^2$ cm$^{-5}$). Right frame: The estimated probability that new discoveries by DES or LSST will enable Fermi's sensitivity to dark matter annihilations to improve by a given value.  Note that the improvement shown here does {\it not} account for increased exposure, but only the inclusion of newly discovered dwarf galaxies (with a total of 10 years of observation, Fermi's sensitivity to dark matter in dwarfs is expected to increase by an additional factor of $\sim$$\sqrt{10/4}\approx 1.6$).}
\label{Figure2}
\end{figure*}

\section{Statistical Approach and Projections}

It is possible to estimate how the discovery of additional dwarf galaxies will improve on Fermi's sensitivity to dark matter in a way that is largely independent of the details of the gamma-ray spectrum and Fermi's instrumental response. In the gaussian limit (applicable for gamma-rays from a large number of dwarfs, at energies at a few tens of GeV or below), we can write:
\begin{equation}
\chi^2 =\frac{(\mathrm{Observed-Background})^2}{(\sqrt{\mathrm{Background}})^2},
\end{equation}
where the above quantities denote the total observed dark matter annihilation signal-plus-background and the total background summed over the combination of the regions surrounding the dwarfs used in a given analysis. To derive the 95\% upper limit on the annihilation cross section, we set $\chi^2=$~3.84. Combining this with Eq.~\ref{flux}, we obtain:
\begin{equation}
\sigv \lsim \frac{8\pi m^2_{\rm DM} (3.84 \, \sum_i B_i)^{1/2}}{A_{\rm eff} \,t^{1/2} \,N_{\gamma} \, (\sum_i J_i)} ,
\end{equation}
where $A_{\rm eff}$ is the effective area of Fermi, $t$ is the duration of the observation, and $B_i$ is the rate of background events in the direction of dwarf, $i$.\footnote{Although we characterize the background for each galaxy using a simple Galactic latitude-dependent model based on data from Fermi, in practice the background can be estimated using more sophisticated diffuse models, or by directly studying the gamma-ray emission observed from the region surrounding the dwarf in question.}

If DES (or LSST) discovers any dwarf galaxies with large $J$-factors, Fermi's sensitivity to dark matter annihilation will be strengthened. In particular, making use of $(N-15)$ new dwarf galaxies discovered by DES or LSST, Fermi's sensitivity to the dark matter annihilation cross section is predicted to improve by a factor given by:
\begin{equation}
\frac{\sigv_{\rm old}}{\sigv_{\rm new}} \simeq \sqrt{\frac{t_{\rm new}}{t_{\rm old}}} \left[ \frac{\sum_{i=1}^{N} J_i}{(\sum_{i=1}^N{B_i})^{1/2}}\right]  \, \left[ \frac{\sum_{i=1}^{15} J_i}{(\sum_{i=1}^{15}{B_i})^{1/2}}\right]^{-1}  
\label{ratio}
\end{equation}

Note that in performing this summation, we include only those newly
discovered dwarfs with large enough $J$-factors to improve upon the
limit. Dwarfs with lesser $J$-factors and/or with large expected
backgrounds that would diminish the overall sensitivity are not
utilized.
By scaling this estimated sensitivity to that already presented by
the Fermi LAT Collaboration, we can present our results in a form that is
approximately independent of quantities such as the particular choice
of the dark matter annihilation channel and Fermi's effective area,
instead depending only the observation time, and on the
$J$-factors and latitudes of the dwarfs to be discovered by DES and/or
LSST. 

Our counts stacking approach represents a considerable simplification
relative to the joint likelihood analysis employed by the Fermi
LAT Collaboration, which would be beyond the scope of this study to
implement. However, our method still incorporates information regarding the
distribution of $J$-factors among detected dwarfs by selecting the combination of dwarfs
in each realization which would yield the highest {\it a priori}
signal-to-noise ratio according to Eq. \ref{ratio}. The
$\chi^2$ treatment presented here has greatest fidelty in the
high-counts (i.e., background-dominated regime) relevant for dark
matter masses of $\lesssim$ 300-500 GeV. For larger masses, a Poisson
treatment should be used.

For simplicity, we assume the uncertainties in the relevant $J$-factors to be negligible (e.g., after spectroscopic follow-up). While this is not likely to be entirely realized, as we treat both currently known dwarfs and to-be-discovered dwarfs in this way, this assumption is unlikely to significantly impact our projections for the {\it improvement} in Fermi's sensitivity. 


In most cases, we find that Fermi's sensitivity to dark matter
annihilations in dwarf galaxies is largely determined by the dwarf
with the largest $J$-factor. Among the 15 dwarfs currently used in the
analysis of the Fermi LAT Collaboration, it is the combination of the few
dwarfs with the largest $J$-factors (Segue 1, Ursa Major II, Wilman 1, and Coma Berenices, each of which $J \gsim 10^{19}$ GeV$^2$ cm$^{-5}$)  that dominate the
calculation of the resulting limit.  Any future discoveries of dwarfs
with $J$-factors below a few times $10^{19}$ GeV$^2$ cm$^{-5}$ are
unlikely to impact this limit significantly.  Instead, in most of the
realizations in which Fermi's sensitivity improves significantly as
the result of the discovery of new dwarfs, it is a single dwarf with
an exceptionally large $J$-factor that accounts for the vast majority
of the improvement.  In other words, significant improvements in Fermi's sensitivity to dark matter are possible, but generally rely on the discovery of a nearby satellite, containing too few stars to have been previously identified as a classical dwarf.

With this in mind, we show in the left frame of Fig.~\ref{Figure2} the
probability of DES or LSST discovering at least one dwarf with a
$J$-factor above a given value.  From this, we see that there is
approximately a 13.4\% (54.5\%) chance that DES (LSST) will discover a
dwarf with a larger $J$-factor than any of the currently known
satellites. In the right frame, instead of focusing on the single dwarf with the largest $J$-factor, we show the estimated likelihood of Fermi's sensitivity to dark matter annihilations improving by a given factor. Note that the improvement shown here does {\it not} account for increased exposure, but only to the inclusion of newly discovered dwarf galaxies. 

\section{Discussion}\label{discussion}

To those hoping that the new dwarf galaxies to-be-discovered by DES or LSST are likely to very significantly improve Fermi's sensitivity to annihilating dark matter, the results presented in the previous section may be disappointing. In this section, we briefly discuss the most important assumptions that have gone into our model, and consider how other approaches could potentially lead to more optimistic projections.

First of all, we reiterate that our model is based on the
mass-luminosity distribution of dwarf galaxies presented in
Ref.~\cite{fbb+11}, and normalized such that there are a total of 25 dwarfs that either qualify as ``classical'' or as discoverable by SDSS. While we consider these choices to be
reasonable, it is possible that they lead to a population model of
satellite galaxies that does not precisely correspond to that of the
Milky Way. 
In particular, we note that our model predicts that
the currently observable satellites with the highest $J$-factors are
likely to be classical dwarfs, rather than those discovered by
SDSS. In reality, however, the three known dwarfs with estimated
$J$-factors greater than $10^{19}$ GeV$^2$ cm$^{-5}$ (Segue 1, Ursa
Major II and Wilman I) were each discovered by SDSS (Coma Berenices, also
discovered with SDSS, has a $J$-factor
near this threshold, $J \simeq 10^{19}$ GeV$^2$ cm$^{-5}$). The
predicted probability of SDSS discovering three dwarfs with $J > 10^{19}$
GeV$^2$ cm$^{-5}$ is approximately 13.6\%. So
while such a realization is not wildly unlikely, it may be indicative
that our model underestimates the number of ultra-faint satellites
with large $J$-factors. 

As an alternative approach to that adopted in
our model, we could have normalized each simulated system such that
SDSS would have been able to discover three or four dwarfs with $J >
10^{19}$ GeV$^2$ cm$^{-5}$. This would increase the number of high
$J$-factor dwarfs predicted to be discovered by DES and LSST by a not
insignificant factor of $\sim$2-3.

According to our model, a fundamental limitation to future
  sensitivity gains is that ultra-faint dwarfs ($\mathrm{M_v}>-8.9$) with $J$-factors similar to or larger than any of
the currently known dwarfs are rare, and ultra-faint dwarfs with $J > 10^{20}$ GeV$^2$ cm$^{-5}$ are
almost non-existant. However, the tail of high $J$-factor
  ultra-faint galaxies might be more prominent in reality than accounted
  for in our model, based on the discussion above. Of course, it is also possible that the SDSS
footprint contains a fortuitously large number of high J-factor
ultra-faint satellites. Estimation techniques that treat the SDSS
sample as perfectly representative are necessarily blind to this
possibility, which motivates simulation-based methods as an
important complementary approach.

\begin{figure}[t!]
\includegraphics[width=3.45in]{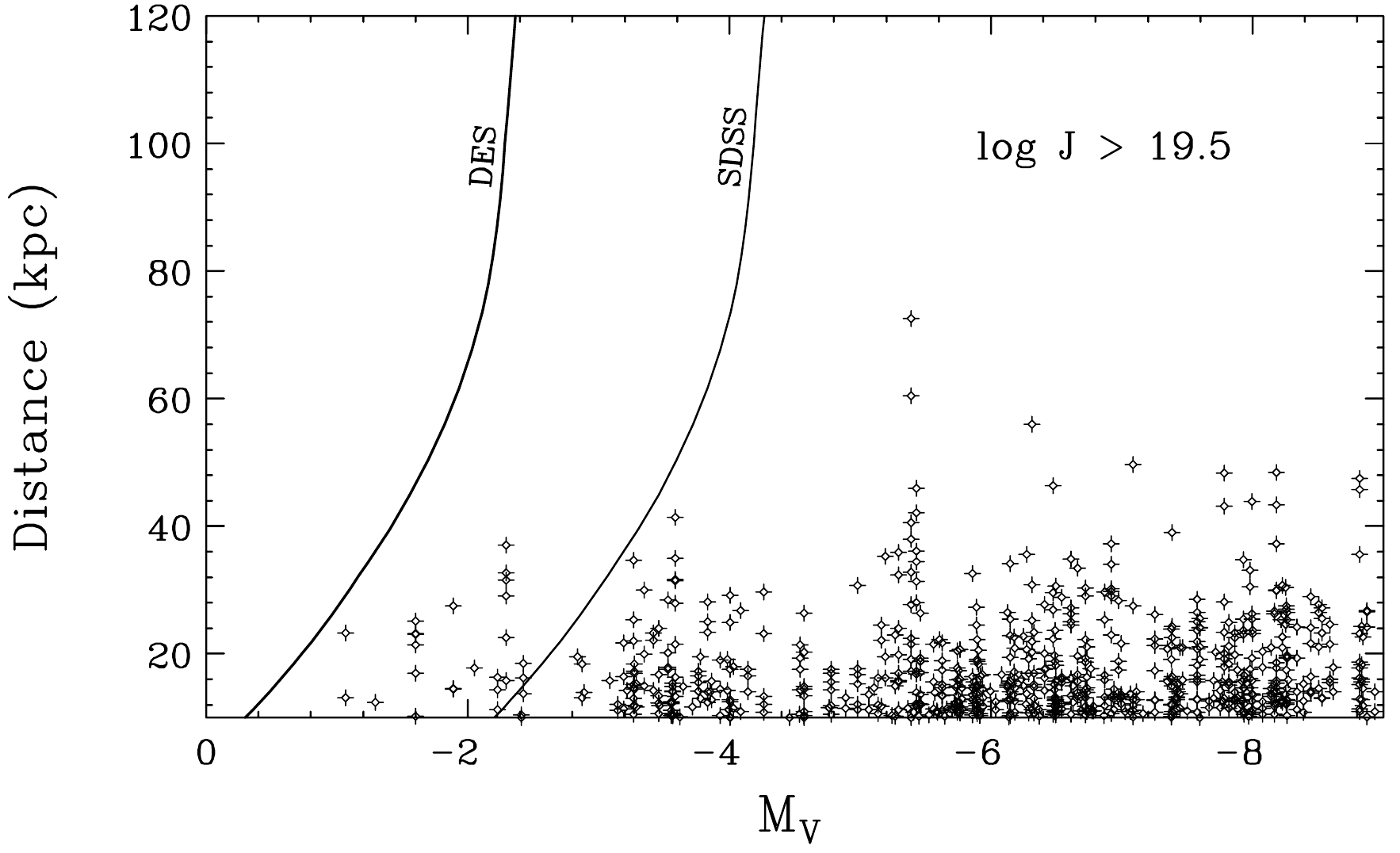}
\caption{The distance to and V-band absolute magnitudes of a random
  sample of dwarf galaxies in our model with $J$-factors larger than
  $10^{19.5}$ GeV$^2$ cm$^{-5}$. Also shown are the approximate thresholds for SDSS or DES to detect a given dwarf galaxy. Note that in our model, all dwarfs with such large $J$-factors are relatively nearby, and few (none) are too faint to be detected by SDSS (DES).}
\label{scatter}
\end{figure}

Our conclusions intrinsically, and not insignificantly, depend on the
luminosity function of dwarf galaxies, which is not currently well constrained observationally at
the faintest luminosities~\cite{Koposov:2007ni}. To illustrate this
dependence, we show in Fig.~\ref{scatter} the distances and V-band
magnitudes for a random sample of dwarfs with $J$-factors larger than
$10^{19.5}$ GeV$^2$ cm$^{-5}$ (the largest value of the currently
known dwarfs), as predicted in our model. This figure illustrates two
key features. Firstly, almost all dwarfs with such large $J$-factors are
located relatively nearby, within $\sim60$ kpc of the Solar
System. And secondly, very few of such dwarfs are too faint to have
been detected by SDSS, and none will be missed by DES or LSST (if
within their fields-of-view).
If we had instead considered a model with a luminosity function
predicting a much larger number of dwarfs with magnitudes fainter than
$M_V\sim-3$, the prospects for DES and LSST could be improved. The model
proposed in Ref.~\cite{tbs+08}, for example, predicts a sharp increase
in the number of dwarfs fainter than $M_V\sim-4$, leading one to
expect more discoveries of high $J$-factor dwarfs by DES and LSST than
is predicted in the model we have used in this study, and thus to more
favorable predictions for the future sensitivity of Fermi to annihilating dark matter. 

\section{Summary}

Using the semi-analytic model of Ref.~\cite{fbb+11}, we have created a large sample of
the satellite populations around Milky Way-like galaxies, and have
used these results to project how future discoveries of dwarf galaxies
by DES and LSST are likely to impact the sensitivity of the Fermi
Gamma-Ray Space Telescope to annihilating dark matter. We find that
the expectations for such improvements are modest, with little chance
that future surveys will increase Fermi's sensitivity to dark matter
by more than a factor of $\sim$2-4. From this perspective, the prospects for improving
Fermi's sensitivity to dark matter annihilating in dwarf galaxies
largely rely on continued observation (i.e., greater exposure)
and from tightening the dynamical constraints on the currently known
dwarfs because the ``best'' targets would probably have already been found.

We caution that these conclusions are based on one set
of modeling choices for the Milky Way satellite population and that more
optimistic forecasts based on different approaches are possible. In
particular, the mock satellite populations considered here seem to be deficient in ultra-faint
galaxies with high $J$-factors, which in reality substantially
strengthen the current limits derived from Fermi observations. 
Further investigations into the luminosity function, radial distribution, and
dark matter halos of the faintest Milky Way companions are
needed to more precisely predict how much DES and LSST will help to improve dark matter
annihilation constraints.

After the initial submission of this work, we became aware of the forecasts of Ref.~\cite{Hargis:2014predictions}. 
Our predictions for the number of dwarf galaxies detectable by DES and LSST are in good agreement with that of Ref.~\cite{Hargis:2014predictions} when considering systems more luminous than $M_V\sim-2.7$, however, the authors of Ref.~\cite{Hargis:2014predictions} separately consider a large number ($\sim100$) of potentially detectable ``hyperfaint'' dwarfs at luminosities $M_V>-2.7$, mostly located beyond 30~kpc. 
The existence of such a population might further enhance the outlook for indirect dark matter searches depending on the properties of the subhalos which host them.

\bigskip
\bigskip

{\it Acknowledgements}: 
Two anonymous referees provided helpful feedback which improved this manuscript.
We would like to thank Andreea Font, Carlos Frenk, and the other
authors of Ref.~\cite{fbb+11} for generously providing us with the
results from their semi-analytic model describing the formation and
evolution of Milky Way satellite galaxies. We would also like to thank Alex Drlica-Wagner, Louie
Strigari, and Johann Cohen-Tanugi for thoughtful comments.  This work has been supported by the US
Department of Energy under Grant No. DE-FG02- 90ER-40560, by the Kavli
Institute for Cosmological Physics at the University of Chicago
through grant NSF PHY-1125897 and an endowment from the Kavli
Foundation and its founder Fred Kavli, and by the National Science
Foundation under Grant No. PHYS-1066293, as well as the Aspen Center for Physics.

\bibliography{DES_citations2.bib}

\end{document}